\documentclass[12pt]{revtex4}
\usepackage{epsfig}
\usepackage{graphicx}
\usepackage{natbib}
\newcommand{\rr}{\mbox{\boldmath $r$}}
\newcommand{\kk}{\mbox{\boldmath $k$}}

\newcommand{\sh}{Schr\"odinger equation}
\newcommand\lt{\left}
\newcommand\rt{\right}
\newcommand\fr{\frac}
\newcommand{\eref}[1]{Eq. (\ref{#1})}
\newcommand{\qm}{quantum mechanics}

\begin{document}

\begin{center}
{\bf Comment On ``On observation of neutron quantum states in the
Earth's gravitational field''~\cite{v}}

V.K.Ignatovich

\end{center}
\bigskip

Criticism in~\cite{v} of experiments~\cite{v1} is not valid. It is
based on misunderstanding of difference between classical and
quantal behavior of particles in classical gravitational
potential. I decided to write this comment because my name is
mentioned in Acknowledgement of~\cite{v}, so the impression can
appear that I approve the interpretation of gravity experiments
given in~\cite{v}. Below I present my vision of ideas of the
experiments made in~\cite{v1}, and make some critical remarks on
the content of the paper~\cite{v}. The more detailed theory can be
found in the papers~\cite{o6,eu} that are referenced in~\cite{v}.

A neutron motion along a horizontal infinitely thick mirror
($z<0$) with the ideal plain interface at $z=0$ and with account
of the Earth gravity field is described by the stationary \sh
\begin{equation}\label{1}
\lt[\fr{\hbar^2}{2m}\lt(\fr{\partial^2}{\partial
x^2}+\fr{\partial^2}{\partial y^2}+\fr{\partial^2}{\partial
z^2}-u\Theta(z<0)\rt)-mgz-E\rt]\Psi(\rr)=0,
\end{equation}
where $E$ is the total energy of the neutron, $m$ is its mass, $g$
is the free fall gravity acceleration, $u=4\pi N_0b$ describes
neutron matter interaction with the horizontal mirror, $b$ is the
coherent amplitude of the neutron scattering from the matter
atoms, $N_0$ is atomic density, and $\Theta(z<0)$ is a step
function equal to unity for $z<0$ and to zero otherwise. After
division of this equation by $\hbar^2/2m$ we reduce it to the form
\begin{equation}\label{2}
\lt[\fr{\partial^2}{\partial x^2}+\fr{\partial^2}{\partial
y^2}+\fr{\partial^2}{\partial
z^2}-u\Theta(z<0)-az-k^2\rt]\Psi(\rr)=0,
\end{equation}
where $a=2m^2g/\hbar^2$ and $k^2=2mE/\hbar^2$.

The variables $x$, $y$, $z$ can be separated. Therefore solution
can be represented as
\begin{equation}\label{3}
\Psi(\rr)=\psi_n(z)\exp(i\kk_\|\rr_\|),
\end{equation}
where $\rr_\|=(x,y,0)$ is two-dimensional radius vector parallel
to the mirror surface and $\psi_n(z)$ is an eigen solution of the
one dimensional equation
\begin{equation}\label{4}
\lt[d^2/dz^2-u\Theta(z<0)-az\Theta(z>0)-k_n^2\rt]\psi_n(z)=0
\end{equation}
with eigen value denoted here by $k_n^2$.  In this equation we
introduced the step function $\Theta(z>0)$ to avoid consideration
of neutron tunnelling through the floor. The tunnelling transforms
the bound states to resonant ones.

The wave vector $\kk_\|$ in \eref{3} is parallel to the mirror
surface and according to \eref{2} and (\ref{4}) its length is
$k_\|=\sqrt{k^2-k_n^2}$. The neutron propagation is free along the
mirror surface and is bound or, in other words, quantized along
its normal, which is parallel to the vertical $z$-axis. The
separation of variables is natural and it does not depend on how
large is $k_\|^2$ comparing to $k_n^2$. We do not denote $\kk_\|$
as $\kk_{n\|}$ for not to give an impression that the motion
parallel to the mirror surface is also quantized. It is not
quantized because there is no potential in the horizontal
direction and $k^2$ can acquire an arbitrary value. However, of
course, for a given $k^2$ different modes $n$ propagate with
different $k_\|$. In experiments~\cite{v1} it was not essential,
because there was $k^2\gg k_n^2$.

We derived all that in details to show that the second of the two
below sentences of the abstract in~\cite{v} is not correct:
\begin{quote}
{\it The Airy functions describe the quantum bouncer (QB), the
concept of which is subject to theoretical study of toy 1D models
of gravitationally bound particles in nonrelativistic quantum
mechanics (QM). This is essentially different from the 3D
nonstationary QM object, ''the running QB,'' investigated in the
experiment.}
\end{quote}

If the horizontal mirror has some static imperfections then for a
given energy $E$, i.e. given value of $k$, the general neutron
wave function because of elastic scattering becomes a
superposition of modes with different $k_n^2$ and different wave
vectors $\kk_{\|n,\phi}$ of the length
$|\kk_{\|n,\phi}|=\sqrt{k^2-k_n^2}$:
\begin{equation}\label{33}
\Psi(\rr,t)=\exp(-iEt/\hbar)\lt[\psi_{n_0}(z)\exp(i\kk_{\|n_0}\rr_\|)+\sum\limits_{n=1}^\infty\int
d\phi a(n,\phi)\psi_n(z)\exp(i\kk_{\|n,\phi}\rr_\|)\rt].
\end{equation}
Here we separated an initial state in the mode $n_0$ as an
incident one. All the coefficients $a(n,\phi)$ ($\phi$ is
azimuthal angle with respect to the initial wave vector
$\kk_{\|n_0}$) can be calculated, say, by perturbation theory. If
imperfections are not static, say mechanical vibration of the
mirror position or elastic waves, then the \sh\ becomes
nonstationary and scattering can be inelastic.

The idea of the experiments~\cite{v1} criticized in~\cite{v} is
very simple. Because of quantization along $z$ axis propagation of
neutrons in $n$-th mode along the horizontal mirror is analogous
to propagation of waves along a waveguide of width $z_n$, where
$z_n$ is the point, after which $\psi_n(z)$ described by Airy
functions starts to decay exponentially. The larger is $n$, the
larger is $z_n$.

If one puts at some height $z_a$ above the reflecting horizontal
mirror an additional horizontal plate with rough lower surface and
complex potential the \sh\ changes. The scattering (their role is
investigated in~\cite{ad}) and absorption (also the tunnelling
through the floor plate) both lead to losses, which can be
described by an imaginary potential $iv$, and equation for a
quantum state $n$ formed at the entrance point satisfies the
equation
\begin{equation}\label{4a}
\lt[d^2/dz^2-u\Theta(z<0)+iv\Theta(z>z_a)-az\Theta(z>0)-K_n^2\rt]\psi_n(z)=0.
\end{equation}
With such a potential the eigen value $K_n^2$ becomes a complex
number: $K_n^2=k'^2_n-ik''^2_n$. From energy conservation we
immediately find that the propagation wave number $k_\|$ along the
plates for every mode also becomes the complex number:
\begin{equation}\label{6}
k_\|=\sqrt{k^2-K_n^2}=\sqrt{k^2-k'^2_n+ik''^2_n}=k'_\|+ik''_{\|n}.
\end{equation}
Therefore the wave of $n$-th mode propagating between the plates
decays proportionally to $\exp(-k''_{n\|}L)$, where $L$ is the
neutron path along the plates. For all $z_n\ge z_a$ absorption is
high, and for $z_n\ll z_a$ the absorption is low, because the Airy
functions exponentially decay at $z_n<z<z_a$. Therefore the gap
between two horizontal plates is a filter, which transmits only
modes with $z_n<z_a$. If $z_a<z_1$ no neutrons are transmitted
through such a filter. The zero transmission at $0<z_a<z_1$ is the
indication of quantization in $z$ direction.

Quantization can be observed even without gravity. The neutron
spectrum between two ideal mirrors with identical potential
barrier $u$, separated by a distance $z_a$ contains quantized and
non quantized parts. The quantized states propagate along plates.
Non quantized states correspond to neutrons penetrating the
plates. If the plates are sufficiently thick and long, and their
potential contains an imaginary part due to absorption or
scattering, the neutrons in non bound states are absorbed. So the
neutrons can be transmitted along the gap between plates only, if
they are in bound states. The bound state levels $E_n=k_n^2/2$ are
determined from the equation~\cite{igu} $R^2\exp(2ik_nz_a)=1$,
where $R$ is reflection amplitude from one of the plate. The
minimal distance, at which the bound state exists, is equal to
$l_1=\pi/\sqrt{u}$, which is of the order of 50 nm. Therefore at
$z_a<l_1$ transmission is zero, and at $z_a=l_1$ there must be a
step in transmission curve.

The gravity level is observed at $z_a=z_1\sim10\mu$ which is 200
times larger than $l_1$. At such distances there are a lot of
bound levels between the plates, and without gravity transmission
would increase linearly in the range $l_1<z_a<z_1$. Of course the
roughnesses and imperfections degrade the quantum step on
transmission curve because after $z_a>l_1$ without gravity, or
$z_a>z_1$ with gravity there is exponential attenuation
$\exp(-2k''_{n\|}L)$ of neutron flux along the plates, however, it
is important to note that these imperfections do not increase
transmission below $z_1$ if not to take into account that they
smear the value of $z_1$ itself.

It is necessary to comment the blue curve in Fig.4 of~\cite{eu}.
In the legend of the insert of the figure it is said that this
curve presents transmission calculated quantum mechanically, but
without gravity. This is a misleading. If we consider an ideal
situation when only gravity is switched off, then in this
situation the optical potentials of the mirrors must be considered
unchanged. In Fig.4 of~\cite{eu} the blue curve is calculated for
a different case: the case when gravity is switched off, but
potentials of the mirrors become infinitely large. So this blue
curve does not correspond to the absence of gravity.

To explain what did they do, let's look, how the losses due to
roughnesses on the surface of the upper mirror are calculated
in~\cite{eu}. It was supposed that we can find channelling wave
function $\psi_n(z)$ between two ideal mirrors in presence of
gravity. This wave function is expressed via Airy functions. If
distance between ideal mirrors is $l$, and height of roughnesses
on the upper mirror is $2\sigma$ then probability of scattering on
roughnesses is estimated as~\cite{eu}
$$\Gamma=\alpha\int\limits_{l-2\sigma}^ldz|\psi_n(z)|^2,
\eqno(9)$$ where $\alpha$ is some loss parameter, which depends on
character of roughnesses. The density of neutrons $|\psi_n(z)|^2$
near roughnesses is determined by Airy functions. The red curve
in~\cite{eu} was calculated with such loss coefficient.

The blue curve in fig.4 of~\cite{eu} is calculated with the wave
function~\cite{eu}
$$\psi_n(z)=\sqrt{\fr2l}\sin\lt(\fr{\pi nz}l\rt).\eqno(21)$$
I.e. this calculation is not related to the case without gravity.
It is related to the case with gravity like the red curve, but
neutron density near roughnesses is approximated by functions
(\cite{eu}21) instead of Airy functions. So it is an approximation
for the same case with gravity. The choice of the function
(\cite{eu}21) is based on assumption that the potential of the
mirrors is infinite. Since it is not true, the choice is voluntary
one and is not related to absence of gravity.

However the blue curve is really some defect of the
paper~\cite{eu}. If the losses in the range $l_1<z_a<z_1$
calculated with the help of rigorous theory without gravity would
give transmission below the background, i.e. unmeasurable in the
experiment, then we cannot be sure that the start of count rate at
10 $\mu$ is really due to the gravity level. The higher losses at
$z_a>z_1$ for blue curve than for red one are not so much
spectacular and can be attributed to some normalization effect.
Nevertheless the criticism of~\cite{v} is aimed not at this defect but at
all the idea of the experiment, and such a criticism is wrong.

The green curve in fig.4 of~\cite{eu}, which shows transmission
when the rough mirror is at bottom instead of ceiling shows
nothing new. It corresponds only to higher losses, which now can
be calculated with formula like
$$\Gamma=\alpha\int\limits_{0}^{2\sigma}dz|\psi_n(z)|^2,
$$
instead of (\cite{eu}9). These losses are higher because
$|\psi_n(0)|^2>|\psi_n(l)|^2$. The only sign of quantization in
gravity is the absence of transmission at $l<\sim10\,\mu$.

 The factor $\exp(-2k''_{n\|}l)$, or $\exp(-\Gamma t)$
in (~\cite{eu}8), where $t=L/v$ and $\Gamma=2k''_{n\|}v$, is the
result of quantization and absorption of quantized states. It can
be estimated by Eq. (3) of~\cite{v}, but not replaced. The author of~\cite{v}
replaced it by his classical expression (~\cite{v}3) and
lost any relation to \qm. His claim (p. 5 before section B)
\begin{quote}
{\it If one doubts our predictions, arguments from rigorous
numerical studies of nonstationary dynamics inside the slit must
be given.}
\end{quote}
is incorrect, because the problem of bound states is purely
stationary, and propagation along the channel is not a classical
bouncing of a point particle between up and down mirrors.

Regrettably there are almost no formula given in~\cite{v}
therefore we have to analyze only words. We will do it by quoting
some parts of conclusion~\cite{v}.
\begin{enumerate}
    \item
According to~\cite{v} the authors of~\cite{v1} failed to see
quantization of the ''running quantum bouncer''
\begin{quote}
{\it because the experiment methodology was not originally
formulated in QM rigor terms.}
\end{quote}
Above derivation shows that it is not true.
    \item The next claim is not founded:
\begin{quote}
{\it We criticize both the model and the measurement method as
being inadequate to the claimed objective of the experiment.}
\end{quote}
    \item The next sentence is surprising:
\begin{quote}
{\it The authors do not attempt to justify their methodology by
conducting a detailed investigation of the problem (Monte Carlo
simulation, for example) as a necessary part of the work.}
\end{quote}
It is not explained how the Monte Carlo method can be used for
solution of the stationary \sh.
    \item The following sentence
\begin{quote}
{\it We also note that the authors do not attempt to gain into a
deeper insight of the QB concept, first of all, the problems of
its definition, existence, and observation. The matter is that
physics of layer of quantum bouncers on a surface of perfect
mirror embraces many aspects well beyond the Airy equation and a
neutron guide problem [23].}
\end{quote}
contains a reference to my work and I cannot understand what does
this sentence mean.
    \item The next claim
\begin{quote}
{\it that the authors' claim, that neutron quantum levels in the
gravitational field of Earth are observed for the first time, is
neither theoretically nor experimentally substantiated.}
\end{quote}
is absolutely not true.
\end{enumerate}

We would like to finish our comment by citation of an experimental
proposal in~\cite{v}:
\begin{quote}
{\it the experiment setup in the quantum regime must be arranged
to measure directly, by definition, a probability of finding a
neutron in the space volume $dV(z)= \Delta x\Delta ydz$ above the
mirror behind the slit. It can be realized, for example, with the
use of microscopic detectors sensitive to neutron wave properties
that is, in a coordinate system comoving with the neutron bouncer
in the $x$ direction. The main point is that the observable QM
object should be the QB rather than ''a running QB'' to allow
standing wave conditions to be satisfied. If so, the transverse
(Airy) mode would be the main quantum mode with energy eigenvalues
being physically meaningful.}
\end{quote}
and we leave it to readers to decide how feasible is it.

\section{History of submission}

I had a correspondence with the author of~\cite{v} prior his publication. I
tried to explain him the essence of the experiment, but failed.
After I saw~\cite{v} I proposed to the author to submit an errata.
He refused then I said that I will submit my comment on his paper.
The comment was submitted to Phys.Rev.D on 18.06 of 2010. One
referees tried to defend paper~\cite{v}, the second one agreed
that the paper~\cite{v} is wrong, but tried to find defects in my
paper. I corrected some points according to his criticism, however
he recommended to reject my comment because I did not criticized
enough the blue curve of Fig.4 in~\cite{eu}. So finally the
editors rejected my comment. I appealed, but appellation system of
APS is wrong. There was a member of the editorial board who wrote
that he concurred with the second referee, and on 8 December of
2010 my appellation was declined. I have no opportunity to present
here the complete referee reports and the name of the editorial
board member who declined my appeal, because of restrictive ArXiv
policy. In this submission to ArXiv I added a paragraph started
with the words ``However the blue curve is really some defect''.
All the other text is not changed. The readers can see how hard
the Phys.Rev.D defended their incompetence, which is demonstrated
by publication of~\cite{v}. Later I will provide the internet
address, where all the referee reports can be found.


\begin{thebibliography}{9}
\bibitem{v}
Anatoli Andrei Vankov. On observation of neutron quantum states in
the Earth's gravitational field. Phys.Rev.D {\bf 81} 052008
(2010).
\bibitem{v1}
V.V. Nesvizhevsky and K.V. Protasov, Quantum States of Neutrons in
the Earth's Gravitational Field: State of the Art, Applications,
Perspectives, Trends in Quantum Gravity Research, edited by D. C.
Moore (Nova Science Publishers, New York, 2006), pp. 65-107.
\bibitem{o6}
 A. E. Meyerovich, V. V. Nesvizhevsky, Gravitational quantum states of neutrons in a rough
 waveguide. Phys. Rev. A {\bf73}, 063616 (2006)
\bibitem{eu}
A.Westphal1, H.Abele, S. Bae$\beta$ler, V.V. Nesvizhevsky,
K.V.Protasov, A.Y. Voronin, A quantum mechanical description of
the experiment on the observation of gravitationally bound states.
Eur. Phys. J. C {\bf51} 367-375 (2007).
\bibitem{ad}
R. Adhikari, Y. Cheng, and A. E. Meyerovich, Quantum size effect
and biased diffusion of gravitationally bound neutrons in a rough
waveguide. Phys. Rev. A {\bf75}, 063613 (2006)
\bibitem{igu}
Vladimir K. Ignatovich, Masahiko Utsuro, {\it Handbook on Neutron
Optics}, Wiley-VCN Verlag GmbH, \& Co. KGaA, 2009.
\end{thebibliography}
\end{document}